\documentclass[twoside]{article}

\usepackage[accepted]{aistats2025}
\usepackage{cite}
\usepackage{balance}
\usepackage{hyperref}
\usepackage{url}
\usepackage{amsmath,amssymb,amsfonts}
\usepackage{graphicx}
\usepackage{textcomp}
\usepackage{xcolor}
\usepackage{amsthm}
\usepackage{mathrsfs}
\usepackage{subcaption}
\usepackage{algorithm}
\usepackage{algorithmic}
\usepackage{placeins}
\usepackage{balance}

\usepackage{pifont}
\usepackage{comment}
\usepackage{tikz}

\newtheorem{proposition}{Proposition}
\newtheorem{lemma}{Lemma}

\DeclareMathOperator*{\argmin}{arg\,min}

\def\BibTeX{{\rm B\kern-.05em{\sc i\kern-.025em b}\kern-.08em
    T\kern-.1667em\lower.7ex\hbox{E}\kern-.125emX}}

\begin{document}

%

%

\twocolumn[

\aistatstitle{Statistically Valid Information Bottleneck via Multiple Hypothesis Testing}

\aistatsauthor{Amirmohammad Farzaneh \And Osvaldo Simeone}

\aistatsaddress{KCLIP Lab, Centre for Intelligent Information Processing Systems (CIIPS)\\
    Department of Engineering, King's College London, London, UK\\
    Email: \{amirmohammad.farzaneh, osvaldo.simeone\}@kcl.ac.uk} ]

\begin{abstract}
The information bottleneck (IB) problem is a widely studied framework in machine learning for extracting compressed features that are informative for downstream tasks. However, current approaches to solving the IB problem rely on a heuristic tuning of hyperparameters, offering no guarantees that the learned features satisfy information-theoretic constraints. In this work, we introduce a statistically valid solution to this problem, referred to as IB via multiple hypothesis testing (IB-MHT), which ensures that the learned features meet the IB constraints with high probability, regardless of the size of the available dataset. The proposed methodology builds on Pareto testing and learn-then-test (LTT), and it wraps around existing IB solvers to provide statistical guarantees on the IB constraints. We demonstrate the performance of IB-MHT on classical and deterministic IB formulations, including experiments on distillation of language models. The results validate the effectiveness of IB-MHT in outperforming conventional methods in terms of statistical robustness and reliability.
\end{abstract}

\section{Introduction}
\label{sec:intro}

\subsection{Context}

\begin{figure}
    \centering
    \includegraphics[width=0.6\columnwidth]{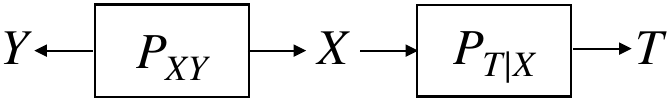}
    \caption{Illustration of the information bottleneck (IB) setup.}
    \label{fig:IB}
\end{figure}

As illustrated in Fig. \ref{fig:IB}, a classical problem in machine learning is extracting a low-dimensional statistic $T$ from an observation $X$ so that $T$ retains sufficient information about a correlated variable $Y$. The more informative $T$ is about $Y$, the more useful $T$ is for downstream inferential tasks targeting variable $Y$.

The information bottleneck (IB) problem, introduced in \cite{tishby2000information}, formalizes this objective by seeking features $T$ of input $X$ such that the mutual information $I(X;T)$ is minimized, while keeping the mutual information $I(T;Y)$ above a user-specified level $\alpha$. This way, the features $T$ remove extraneous information present in $X$ that does not correlate with $Y$, while ensuring that $T$ contains enough information about $Y$ \cite{zaidi2020information}. Specifically, the IB problem for a pair of random variables $(X,Y)\sim P_{XY}$ can be stated as the constrained problem
\begin{align}
\label{eq:IB}
    & \underset{P_{T|X}}{\text{minimize}}\quad I(X;T) \nonumber \\
    &\text{subject to} \quad I(T;Y)\geq \alpha,
\end{align}
where $\alpha\geq 0$ determines the minimum acceptable value of the mutual information $I(T;Y)$, and the minimization is taken over all stochastic mappings $P_{T|X}$. Note that, throughout this article, we focus on the case of variables $X$, $Y$, and $T$ taking values in discrete finite alphabets.

Since its introduction, the IB problem (\ref{eq:IB}) has found its way into numerous applications ranging from clustering \cite{slonim1999agglomerative} to DNN classifiers \cite{achille2018information} and generative models \cite{higgins2017beta}. A common approach to address problem (\ref{eq:IB}) is to introduce a Lagrange multiplier $\lambda>0$ to tackle the unconstrained problem
\begin{equation}
\label{eq:classic_IB}
    \underset{P_{T|X}}{\text{minimize}} \quad  I(X;T)- \lambda I(T;Y).
\end{equation}
As a generalization, reference \cite{strouse2017deterministic} proposed to address the problem
\begin{equation}
\label{eq:deterministic_IB2}
    \underset{P_{T|X}}{\text{minimize}} \quad H(T)-\gamma H(T|X) - \beta I(T;Y),
\end{equation} 
which includes two hyperparameters $\lambda = (\gamma,\beta)$.

In practice, one often only has access to a data set $\mathcal{D}$ of $n$ i.i.d. samples from the joint distribution $P_{XY}$, and not directly to the joint distribution $P_{XY}$. The conventional approach in this case is to tackle problems (\ref{eq:classic_IB}) or (\ref{eq:deterministic_IB2}) by substituting the two mutual informations with empirical estimates based on data set $\mathcal{D}$. However, there is currently no systematic way to choose the hyperparameters $\lambda$ in (\ref{eq:classic_IB}) and (\ref{eq:deterministic_IB2}) so as to satisfy the constraint in (\ref{eq:IB}) \cite{alemi2016deep,tishby2015deep}.

\subsection{Statistically Valid Information Bottleneck}

In this context, this work presents a general hyperparameter optimization (HPO) methodology that wraps around any existing solver for the IB problem to ensure that, when a solution is found, the constraint in (\ref{eq:IB}) is met, irrespective of the size of the data set $\mathcal{D}$. More precisely, given any solver returning a mapping $P^\lambda_{T|X}$ dependent on hyperparameters $\lambda$, whenever it produces an output, the proposed method returns a hyperparameter vector $\lambda$ that is guaranteed to satisfy the relaxed constraint
\begin{equation}
\label{eq:relaxed_constraint}
    \mathrm{Pr}[I(T;Y)\geq \alpha]\geq 1-\delta
\end{equation}
for a given user-defined outage level $0<\delta <1$, where the probability is evaluated over the distribution of the data set $\mathcal{D}$.

We build the proposed approach, termed IB via multiple hypothesis testing (IB-MHT), on Pareto testing \cite{laufer2022efficiently}, an HPO method that provides statistical guarantees on the average risk. Pareto testing in turn leverages learn-then-test (LTT) \cite{angelopoulos2021learn}, which formulates the problem of HPO as an instance of multiple hypothesis testing (MHT).

Accordingly, as illustrated in Fig. \ref{fig:IB_MHT}, IB-MHT first estimates a Pareto frontier on the plane $(I(T;Y), I(X;T))$ based on a portion of the available data $\mathcal{D}$, and then it sequentially tests candidate hyperparameters $\lambda$ in order of decreasing estimated mutual information $I(T;Y)$. By adopting a family-wise error rate (FWER) sequential testing method to identify the final hyperparameter $\lambda^*$, IB-MHT can provably meet the requirement (\ref{eq:relaxed_constraint}), while approximately minimizing the objective in (\ref{eq:IB}).

\subsection{Main Contributions}

The main contributions of this paper are as follows.
\begin{itemize}
    \item \textit{Problem formulation: }We present a probabilistic formulation of the IB problem, which relaxes the constraints of the original formulation into the probabilistic requirement (\ref{eq:relaxed_constraint}) with respect to the available data $\mathcal{D}$.
    \item \textit{Methodology: }We introduce IB-MHT, a hyperparameter selection methodology that wraps around any existing IB problem solvers to return solutions that are guaranteed to satisfy the requirement (\ref{eq:relaxed_constraint}) for discrete random variables.
    \item \textit{Applications: }We detail the application of IB-MHT to existing solvers based on the variational IB \cite{alemi2016deep} and on formulations (\ref{eq:classic_IB}) and (\ref{eq:deterministic_IB2}), including experiments on the problem of distilling language models \cite{zhang2023text}.
\end{itemize}

The rest of this paper is organized as follows. In Section \ref{sec:conventional_IB}, we provide a description of conventional solvers for the IB problem. We introduce IB-MHT in Section \ref{sec:IB-MHT}, and prove that it guarantees the statistical constraint (\ref{eq:relaxed_constraint}). We provide simulation results in Section \ref{sec:simulations}, and conclude the paper in Section \ref{sec:conclusion} and introduce potential future research directions.

\section{Conventional Information Bottleneck Solvers}
\label{sec:conventional_IB}

In this section, we briefly review standard solvers for the IB problem (\ref{eq:IB}). Throughout, we assume that variables $X$, $Y$, and $T$ take values in discrete finite sets $\mathcal{X}$, $\mathcal{Y}$, and $\mathcal{T}$ with respective sizes $|\mathcal{X}|$, $|\mathcal{Y}|$, and $|\mathcal{T}|$.

To this end, assume the availability of a data set $\mathcal{D} =  \left\{(X_i,Y_i)\right\}_{i=1}^n$ with data points drawn i.i.d. from the joint distribution $P_{XY}$. As discussed in Section \ref{sec:intro}, a typical way to address the IB problem (\ref{eq:IB}) is via the unconstrained problems (\ref{eq:classic_IB}) or (\ref{eq:deterministic_IB2}). Conventional solvers address problems (\ref{eq:classic_IB}) or (\ref{eq:deterministic_IB2}) by substituting the mutual informations $I(X;T)$ and $I(T;Y)$ with estimates based on the data set $\mathcal{D}$.

In some very specific cases, such as for doubly symmetric binary sources \cite{zaidi2020information}, the minimization problem can be solved in closed form. However, in practice, the distribution $P_{T|X}$ is modeled as a parametric function such as a deep neural network, and problem (\ref{eq:classic_IB}) or (\ref{eq:deterministic_IB2}) are addressed using gradient-based stochastic optimization strategies. This methodology is known as variational IB (VIB) \cite{alemi2016deep}.

Variational IB thus provides solvers that return mappings $P^\lambda_{T|X}$ dependent on the choice of the hyperparameter vector $\lambda$. For example, the hyperparameter $\lambda$ in (\ref{eq:classic_IB}) dictates the trade-off between compression, as measured by the mutual information $I(X;T)$, and the information $I(T;Y)$ retained by $T$ about $Y$. A smaller $\lambda$ would encourage more compression, and a larger $\lambda$ would prioritize informativeness. Consequently, tuning the value of $\lambda$ is a critical design choice that must be taken into account when solving the IB problem in order to tackle the constraint in (\ref{eq:IB}). However, no systematic procedure is currently known to select the hyperparameters $\lambda$ so as to provably meet the constraint in problem (\ref{eq:IB}).

To the best of our knowledge, there has been no systematic way for optimizing the hyperparameter $\lambda$, even though the need for such methodologies has been identified and discussed \cite{alemi2016deep}. For instance, the simulation results in \cite{shwartz2017opening} are obtained by using a fixed value for $\lambda$. More sophisticated approaches, such as those adopted in \cite{alemi2016deep, tishby2015deep}, solve (\ref{eq:classic_IB}) several times for different values of $\lambda$ by using the same training data set $\mathcal{D}$. Then, the value of $\lambda$ is selected that meets the estimated constraint $\hat{I}(T;Y)\geq \alpha$ while minimizing the estimated mutual information $\hat{I}(X;T)$.

\section{Information Bottleneck via Multiple Hypothesis Testing}
\label{sec:IB-MHT}

In this section, we introduce the proposed IB-MHT approach to address the IB problem (\ref{eq:IB}) for discrete random variables. IB-MHT wraps around any existing solver for the IB problem (\ref{eq:IB}), described in Section \ref{sec:conventional_IB}, and it will be shown to meet the statistical constraint (\ref{eq:relaxed_constraint}).

\subsection{Estimating the Mutual Information}

To start, consider the problem of estimating a mutual information $I(U;V)$ for jointly distributed random variables $(U,V)\sim P_{UV}$ taking values in discrete finite sets $\mathcal{U}$ and $\mathcal{V}$, respectively. To this end, we have access to a data set $\mathcal{D} = \{(U_1,V_1),\ldots, (U_n, V_n)\}$, with samples drawn i.i.d. from the joint distribution $P_{UV}$. A plug-in estimator of the mutual information $I(U;V)$ first evaluates the empirical joint distribution, or histogram, $\hat{P}_{UV}$, and then obtains the estimate
\begin{equation}
\label{eq:I_estimate}
    \hat{I}_\mathcal{D}(U; V) = \sum_{u\in \mathcal{U}} \sum_{v \in \mathcal{V}} \hat{P}_{UV}(u, v) \log \left( \frac{\hat{P}_{UV}(u, v)}{\hat{P}_U(u) \hat{P}_V(v)} \right),
\end{equation}
where $(\hat{P}_U, \hat{P}_V)$ represent the marginal empirical distributions obtained from the joint empirical distribution $\hat{P}_{UV}$.

The following result, proved in \cite{stefani2014confidence}, provides a statistically valid upper bound on the error of the plug-in estimator (\ref{eq:I_estimate}).

\begin{lemma}[\!\!\cite{stefani2014confidence}]
\label{lemma}
    For any probability $0<\epsilon<1$, the estimator (\ref{eq:I_estimate}) satisfies the inequality 
\begin{equation}
\label{eq:confidence}
    \mathrm{Pr}[\hat{I}_\mathcal{D}(U;V)-I(U;V) \leq \Delta I(\theta(\epsilon,n))] \geq 1-\epsilon,
\end{equation}
where
\begin{equation}
\label{eq:theta}
        \theta(\epsilon,n) = \sqrt{\frac{2}{n} \ln \left(\frac{2^{|\mathcal{U}||\mathcal{V}|}-2}{\epsilon}\right)},
    \end{equation}
    and
    \begin{equation}
    \label{eq:delta}
    \Delta I(\theta) =
    \begin{cases}
        \begin{aligned}
            &\frac{\theta}{2} \log [ (|\mathcal{U}||\mathcal{V}| - 1)\\
            &\quad (|\mathcal{U}| - 1)(|\mathcal{V}| - 1) ] \\
            & \quad+ 3h\left( \frac{\theta}{2} \right)
        \end{aligned} & \text{if } \theta \leq 2 - \frac{2}{|\mathcal{U}|}, \\
        \log |\mathcal{U}| & \text{if } \theta > 2 - \frac{2}{|\mathcal{U}|},
    \end{cases}
\end{equation}
with $h(x) = -x\log x -(1-x)\log (1-x)$ being the binary entropy function.
\end{lemma}

\subsection{IB-MHT: IB via Multiple Hypothesis Testing}

IB-MHT wraps around any existing IB solver that returns a hyperparameter-dependent mapping $P^\lambda_{T|X}$. The goal of IB-MHT is to use the data set $\mathcal{D} = \left\{(X_i,Y_i)\right\}_{i=1}^n$, with i.i.d. samples from distribution $P_{XY}$, to return a hyperparameter $\lambda^*$ that approximately minimizes the objective $I(X;T)$ in (\ref{eq:IB}), while guaranteeing the probabilistic constraint (\ref{eq:relaxed_constraint}). To this end, IB-MHT starts with a pre-selected set $\Lambda$ of candidate hyperparameters $\lambda$. The pre-selection can be done using any criterion, as long as it does not use the available data $\mathcal{D}$.

\begin{figure}
    \centering
    \includegraphics[width=\columnwidth]{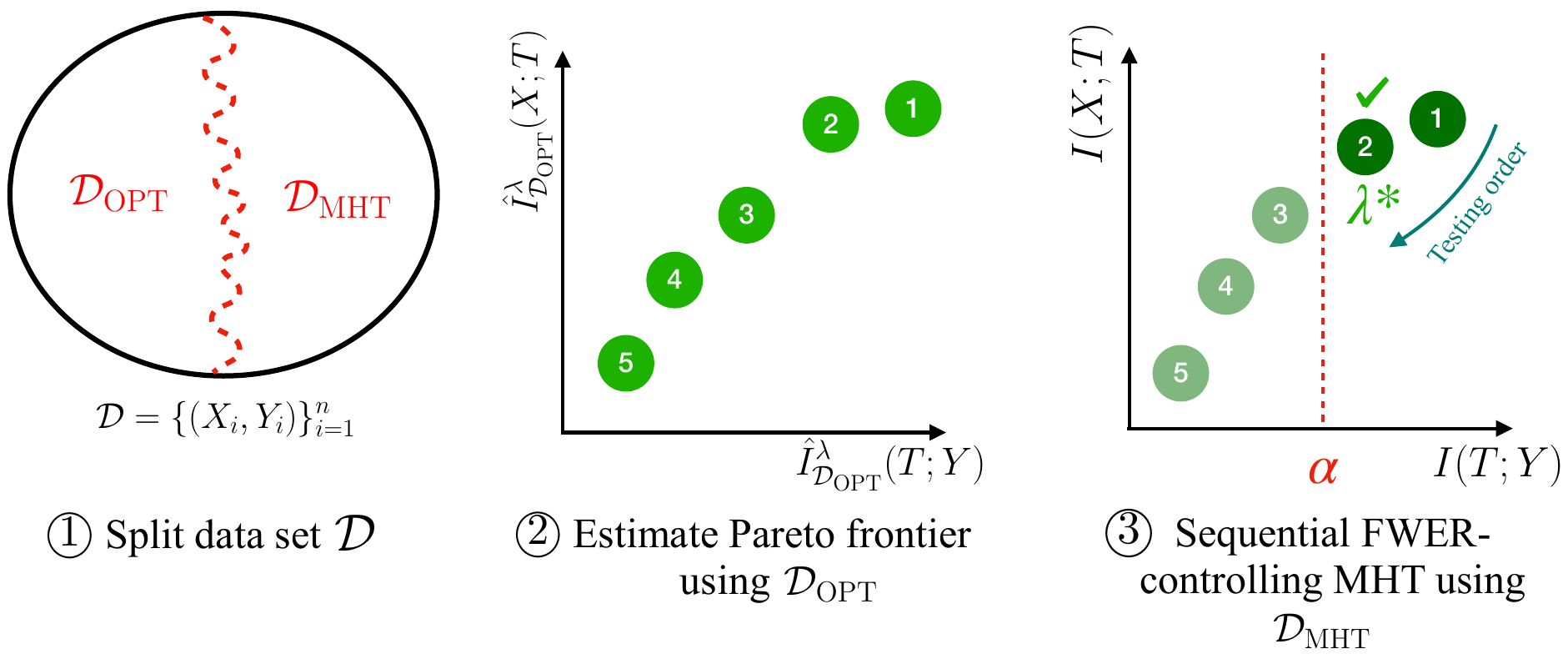}
    \caption{Illustration of the operations of IB-MHT: \textcircled{1} The calibration data set $\mathcal{D}$ is split into two disjoint subsets $\mathcal{D}_\text{OPT}$ and $\mathcal{D}_\text{MHT}$. \textcircled{2} The Pareto frontier in the plane $(I(T;Y),I(X;T))$ is estimated by using the mutual information estimates $\hat{I}^\lambda_{\mathcal{D}_\text{OPT}}(T;Y)$ and $\hat{I}^\lambda_{\mathcal{D}_\text{OPT}}(X;T)$ to obtain the ordered subset $\Lambda_{\text{OPT}}$. \textcircled{3} FST, a sequential FWER-controlling MHT algorithm, is applied to the subset $\Lambda_{\text{OPT}}$ to form the subset $\Lambda_{\text{MHT}}\subseteq\Lambda_{\text{OPT}}$ of hyperparameters $\lambda \in \Lambda_{\text{MHT}}$ that are likely to satisfy the constraint (\ref{eq:relaxed_constraint}). Finally, the hyperparameter $\lambda^*$ is chosen as the vector in $\Lambda_{\text{MHT}}$ that minimizes the estimate $\hat{I}^\lambda_{\mathcal{D}_\text{MHT}}(X;T)$.}
    \label{fig:IB_MHT}
\end{figure}

As illustrated in Fig. \ref{fig:IB_MHT}\textcircled{1}, the proposed IB-MHT follows a two-step procedure, which relies on a split of the available data set $\mathcal{D}$ into two disjoint subsets $\mathcal{D}_{\text{OPT}}$ and $\mathcal{D}_{\text{MHT}}$ of sizes $n_{\text{OPT}}$ and $n_{\text{MHT}}$, respectively, where $n_{\text{OPT}}+n_{\text{MHT}} = n$.

Write as $I^\lambda(X;T)$ and as $I^\lambda(T;Y)$ the ground-truth mutual informations obtained under the joint distribution given by the product of $P_{XY}$ and $P^\lambda_{T|X}$. In the first step (Fig. \ref{fig:IB_MHT}\textcircled{2}), IB-MHT uses the optimization data $\mathcal{D}_{\text{OPT}}$ to find an approximate Pareto front on the plane $(I(T;Y), I(X;T))$, along with the associated subset $\Lambda_\text{OPT}\subseteq \Lambda$ of candidate hyperparameters returning pairs $(I^\lambda(T;Y), I^\lambda(X;T))$ on the Pareto front.

This is done by first obtaining the estimated pairs $(\hat{I}_{\mathcal{D}_\text{OPT}}^\lambda(T;Y),\hat{I}_{\mathcal{D}_\text{OPT}}^\lambda(X;T))$ for all candidate hyperparameter vectors $\lambda \in \Lambda$. Once all such pairs are evaluated, the non-dominated pairs form the estimated Pareto front (green circles in Fig. \ref{fig:IB_MHT}\textcircled{2}). A non-dominated pair $(\hat{I}_\mathcal{D}^\lambda(T;Y), \hat{I}_\mathcal{D}^\lambda(X;T))$ is one for which no other hyperparameter $\lambda' \in \Lambda$ satisfies the inequalities $(\hat{I}^{\lambda'}_\mathcal{D}(T;Y)\geq \hat{I}_\mathcal{D}^\lambda(T;Y), \hat{I}^{\lambda'}_\mathcal{D}(X;T)\leq \hat{I}^\lambda_\mathcal{D}(X;T))$ with at least one inequality being strict.

The second step of IB-MHT (Fig. \ref{fig:IB_MHT}\textcircled{3}) is to apply MHT to the $|\Lambda_{\text{OPT}}|$ null hypotheses
\begin{equation}
\label{eq:hypothesis}
    \mathcal{H}_{\lambda} : I_\lambda(T; Y)<\alpha
\end{equation}
for all hyperparameters $\lambda \in \Lambda_{\text{OPT}}$. The null hypothesis $\mathcal{H}_\lambda$ assumes that hyperparameter $\lambda$ does not meet the constraint in (\ref{eq:IB}). By the definition (\ref{eq:hypothesis}), rejecting the null hypothesis $\mathcal{H}_{\lambda}$ is equivalent to deciding that hyperparameter $\lambda$ satisfies the constraint $I^\lambda(T; Y)\geq \alpha$ in (\ref{eq:IB}).

To this end, IB-MHT lists the hyperparameters in set $\Lambda_\text{OPT}$ in order of decreasing values of the estimate $\hat{I}^\lambda_{\mathcal{D}_\text{OPT}}(T;Y)$, i.e.,
\begin{equation}
    \hat{I}^{\lambda_{(1)}}_{\mathcal{D}_\text{OPT}}(T;Y) \geq \hat{I}^{\lambda_{(2)}}_{\mathcal{D}_\text{OPT}}(T;Y) \geq \ldots \hat{I}^{\lambda_{(|\Lambda_{\text{OPT}}|)}}_{\mathcal{D}_\text{OPT}}(T;Y),
\end{equation}
obtaining the ordering $(\lambda_{(1)}, \ldots, \lambda_{(|\Lambda_{\text{OPT}}|)})$.

To test the hypotheses $\mathcal{H}_\lambda$ for all $\lambda \in \Lambda_\text{OPT}$, IB-MHT uses a sequential family-wise error rate (FWER) controlling algorithm based on the data set $\mathcal{D}_{\text{MHT}}$. Accordingly, as in Pareto testing \cite{laufer2022efficiently}, the hyperparameters in the set $\Lambda_{\text{OPT}}$ are tested in the order $\lambda_{(1)}, \ldots,\lambda_{(|\Lambda_{\text{OPT}}|)}$. At the end of this testing process, to be detailed in the next section, a subset $\Lambda_{\text{MHT}}\subseteq \Lambda_\text{OPT}$ of hyperparameters is selected with the property that, with high probability, the set contains no hyperparameter $\lambda$ that violates the constraint in (\ref{eq:IB}).

Finally, the hyperparameter $\lambda^*$ is selected as the hyperparameter in the set $\Lambda_{\text{MHT}}$ that minimizes the estimate $\hat{I}_{\mathcal{D}_\text{MHT}}^\lambda(X;T)$. Note that if the set $\Lambda_{\text{MHT}}$ is empty, IB-MHT returns an empty set, indicating that IB-MHT cannot guarantee the constraint (\ref{eq:relaxed_constraint}) for any hyperparameter $\lambda$. Algorithm \ref{alg:Pareto} summarizes these steps. The implementation of this algorithm can be found at \href{https://github.com/kclip/IB-MHT}{https://github.com/kclip/IB-MHT}.

\subsection{MHT via Fixed Sequence Testing}

To perform MHT on the set of candidate hyperparameters $\Lambda_{\text{OPT}}$, we first use the following Proposition to form valid $p$-values for all the hypotheses (\ref{eq:hypothesis}). A $p$-value $\hat{p}_\lambda$ for the null hypothesis $\mathcal{H}_\lambda$ is a random variable that satisfies the condition $\mathrm{Pr}[\hat{p}_\lambda \leq u \mid \mathcal{H}_\lambda]\leq u$ for all $u\in [0,1]$.

\begin{proposition}
\label{Proposition::interval_to_pvalue}
    The quantity
    \begin{equation}
        \hat{p}_\lambda= \inf \{\epsilon \in [0,1]: \hat{I}_{\mathcal{D}_{\text{MHT}}}^\lambda(T;Y)-\Delta I(\theta(\epsilon, n)) \leq \alpha \}
    \end{equation}
    is a valid $p$-value for the null hypothesis (\ref{eq:hypothesis}), where $\theta(\epsilon, n)$ and $\Delta I(\theta)$ are defined as in (\ref{eq:theta}) and (\ref{eq:delta}), with $\mathcal{U} = \mathcal{T}$ and $\mathcal{V} = \mathcal{Y}$.

\end{proposition}
\begin{proof}

The validity of the $p$-value $\hat{p}_\lambda$ follows directly from the standard steps \cite[Chapter 9]{rice2007mathematical}
    \begin{align}
        &\mathrm{Pr}_{\mathcal{D}_{\text{MHT}}}[\hat{p}_{\lambda} \leq u \mid \mathcal{H}_{\lambda}] \nonumber \\
        & = \mathrm{Pr}_{\mathcal{D}_{\text{MHT}}}[\hat{I}_{\mathcal{D}_{\text{MHT}}}^\lambda(T;Y)-\Delta I(\theta(u,n))> \alpha \mid I^\lambda(T;Y) < \alpha ] \nonumber \\
        &\leq  \mathrm{Pr}_{\mathcal{D}_{\text{MHT}}}[\hat{I}_{\mathcal{D}_{\text{MHT}}}^\lambda(T;Y)-\Delta I\left(\theta(u,n)\right)>  I^\lambda(T;Y)] \nonumber \\
        &\leq u, 
    \end{align}
    where the probability is evaluated with respect to the source of data set $\mathcal{D}_\text{MHT}$, and the last inequality follows from Lemma \ref{lemma}.
\end{proof}

With the $p$-values in Proposition \ref{Proposition::interval_to_pvalue}, IB-MHT applies fixed sequence testing (FST) \cite{bauer1991multiple} by considering each hyperparameter in set $\Lambda_{\text{OPT}}$ in order starting from $\lambda_{(1)}$, stopping at the first hyperparameter $\lambda_{(j)}$ that does not satisfy the inequality $\hat{p}_{\lambda_{(j)}} \leq \delta$. It then forms the subset $\Lambda_{\text{MHT}}$ as
\begin{equation}
    \Lambda_{\text{MHT}} = \{\lambda_{(1)}, \ldots, \lambda_{(j)}\}.
\end{equation}

\begin{algorithm}
\caption{IB-MHT}
\label{alg:Pareto}
\begin{algorithmic}
    
    \STATE \textbf{Input:} Candidate set $\Lambda$, subsets $\mathcal{D}_{\text{OPT}}$ and $\mathcal{D}_{\text{MHT}}$ from calibration data $\mathcal{D}$

    \STATE \textbf{Output:} Approximate solution $\lambda^*$ to (\ref{eq:IB}) satisfying (\ref{eq:relaxed_constraint})

    \STATE Evaluate the approximate Pareto front $\Lambda_{\text{OPT}}$ using estimates $(\hat{I}_{\mathcal{D}_\text{OPT}}^\lambda(T;Y),\hat{I}_{\mathcal{D}_\text{OPT}}^\lambda(X;T))$
    \STATE Order set $\Lambda_{\text{OPT}}$ according to the values $\hat{I}^{\lambda}_{\mathcal{D}_\text{OPT}}(T;Y)$ from high to low
    \STATE Compute $\hat{p}_\lambda$ for all $\lambda \in \Lambda_{\text{OPT}}$ using Proposition \ref{Proposition::interval_to_pvalue}
    \STATE Apply FST to the $p$-values $\hat{p}_\lambda$ using the ordered set $\Lambda_{\text{OPT}}$ to obtain set $\Lambda_{\text{MHT}}$
    \IF{$\Lambda_{\text{MHT}}$ is not empty}
        \STATE $\lambda^* = \argmin_{\lambda \in \Lambda_{\text{MHT}}}\{\hat{I}_{\mathcal{D}_{\text{MHT}}}^\lambda(X;T)\}$
    \ELSE
        \STATE $\lambda^* = \emptyset$
    \ENDIF

    \RETURN $\lambda^*$
\end{algorithmic}
\end{algorithm}

\subsection{Analysis of IB-MHT}

The hyperparameter $\lambda^*$ returned by Algorithm \ref{alg:Pareto} is guaranteed to meet the constraint (\ref{eq:relaxed_constraint}), as stated in the following proposition.

\begin{proposition}[\!\!{\cite[Proposition 5.1]{laufer2022efficiently}}]
\label{Proposition::gaurantee}
For any $0<\delta <1$, the hyperparameter $\lambda^*$ returned by Algorithm \ref{alg:Pareto} is guaranteed to satisfy the constraint (\ref{eq:relaxed_constraint}).
\end{proposition}

\section{Experiments for Image Representation}
\label{sec:simulations}

\subsection{Problem Setting}

In this section, as in \cite{alemi2016deep}, the training data set consists of 60,000 data points from the binary MNIST training data set. We adopt a neural network model $P^\lambda_{T\mid X}$ trained via VIB based first on objective (\ref{eq:classic_IB}) and then on objective (\ref{eq:deterministic_IB2}). The selection of the hyperparameter $\lambda$ is based on a separate calibration data set $\mathcal{D}$ of 5,000 data points from the binary MNIST test dataset. Note that the data points satisfy the condition of being discrete, as required for IB-MHT to be applicable. For each run of Algorithm \ref{alg:Pareto}, we randomly split data set $\mathcal{D}$ into two disjoint subsets $\mathcal{D}_{\text{OPT}}$ and $\mathcal{D}_{\text{MHT}}$ of size 2,500. Additionally, to test the returned hyperparameters $\lambda^*$, we used an additional 5,000 images from the binary MNIST test data set.

\subsection{Classical IB Problem}
\label{sec:VIB}

Considering first the classical IB problem (\ref{eq:classic_IB}), the initial set of candidate scalar hyperparameters $\Lambda$ contains 100 logarithmically spaced points in the interval $[10^{-4}, 1]$, the outage level is set to $\delta = 0.1$, and the threshold for the constraint (\ref{eq:relaxed_constraint}) is set to $\alpha = 2.28$. 

\begin{figure}[t]
    \centering
    \begin{subfigure}[b]{0.7\columnwidth}
        \centering
        \includegraphics[width=\textwidth]{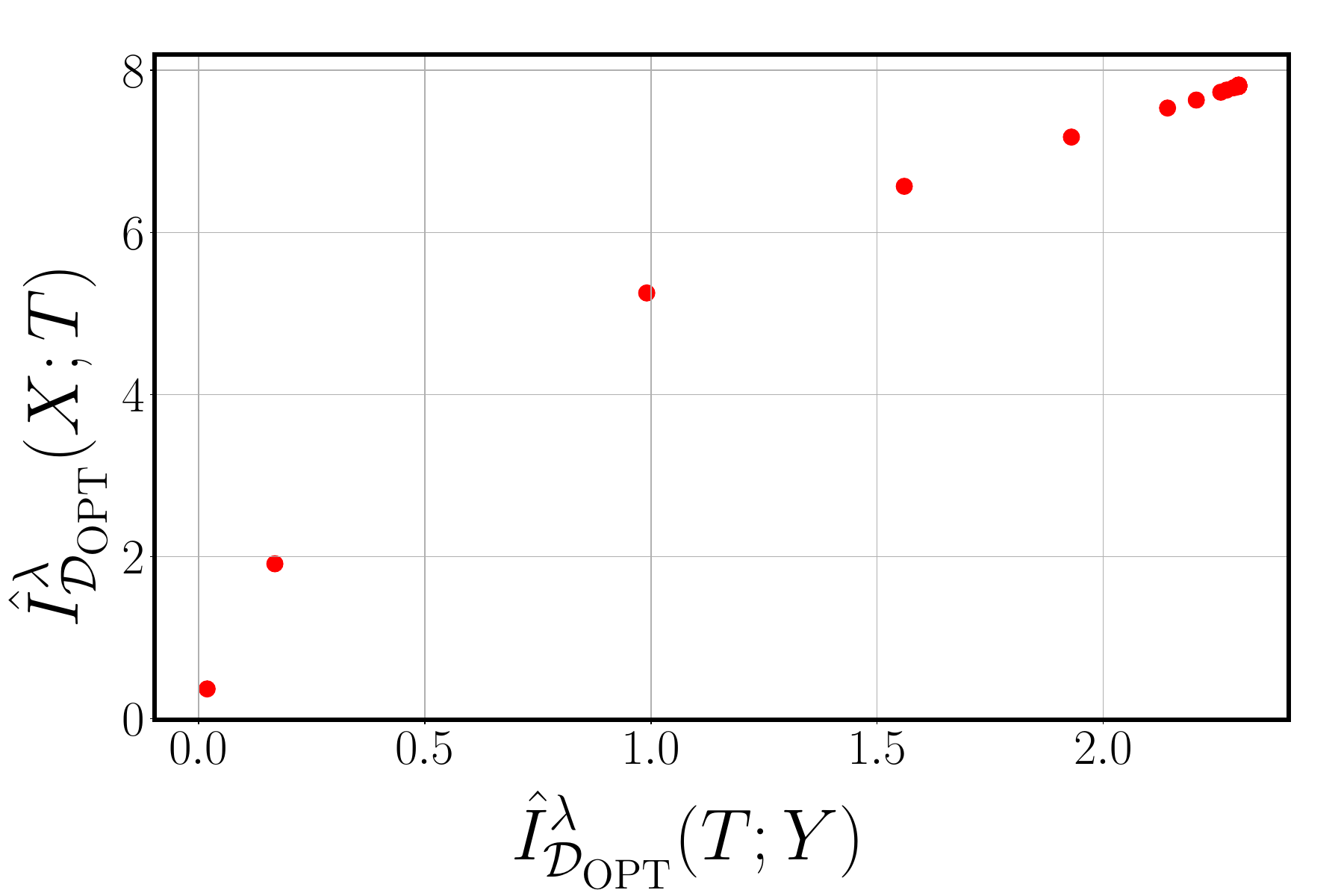}
        \caption{}
    \end{subfigure}
    \begin{subfigure}[b]{0.7\columnwidth}
        \centering
        \includegraphics[width=\textwidth]{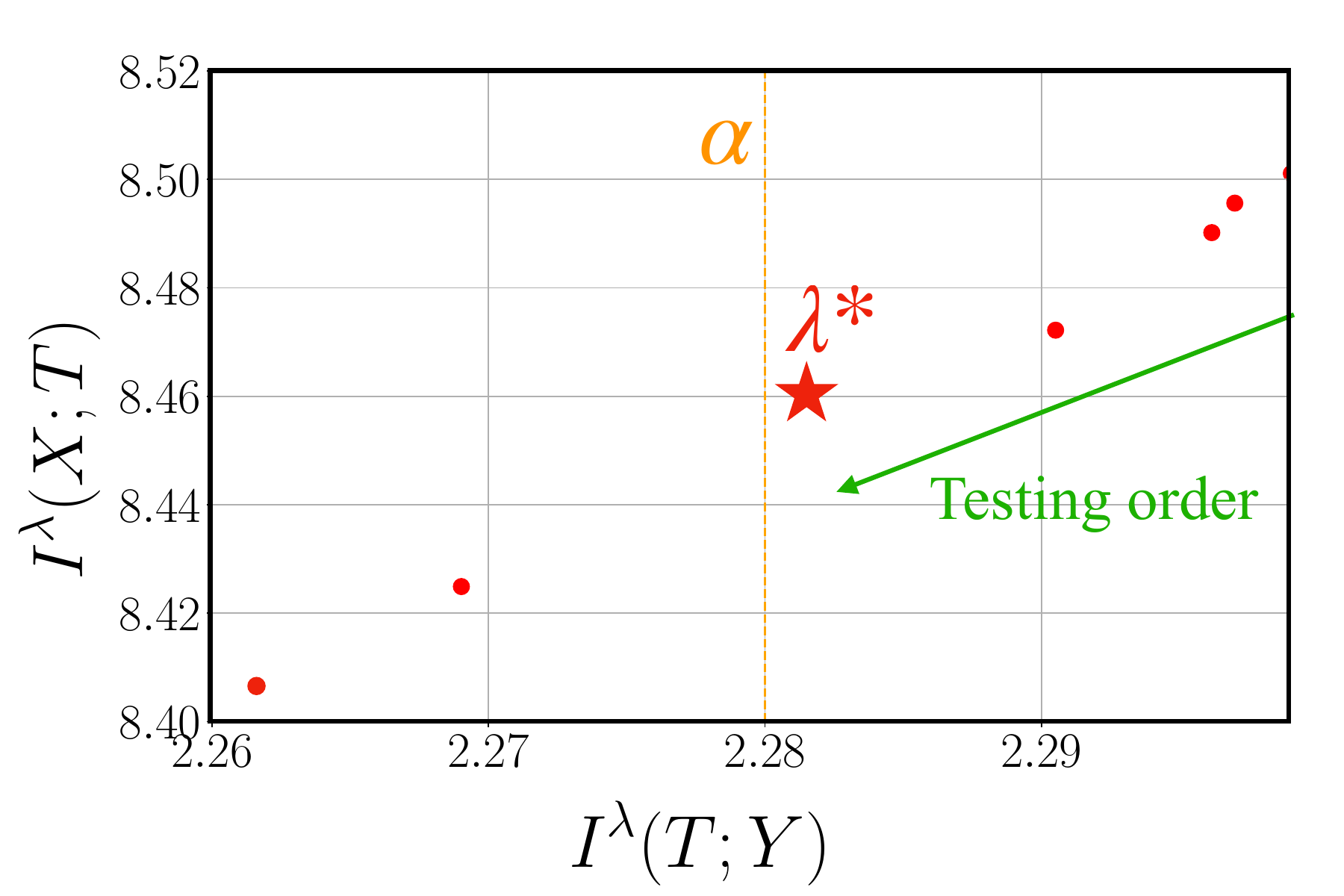}
        \caption{}
    \end{subfigure}
    \caption{Illustration of the operation of IB-MHT for the experiment in Section \ref{sec:simulations}: (a) Estimated Pareto front using the estimated mutual informations $\hat{I}^\lambda_{\mathcal{D}_\text{OPT}}(T;Y)$ and $\hat{I}^\lambda_{\mathcal{D}_\text{OPT}}(X;T)$; (b) Sequential MHT using the estimated mutual informations $\hat{I}^\lambda_{\mathcal{D}_\text{MHT}}(T;Y)$ and $\hat{I}^\lambda_{\mathcal{D}_\text{MHT}}(X;T)$.} 
    \label{fig:steps}
\end{figure}

To start, Fig. \ref{fig:steps} illustrates the operation of IB-MHT in a manner similar to Fig. \ref{fig:IB_MHT}. Specifically, the Pareto front given by the estimates $(\hat{I}_{\mathcal{D}_\text{OPT}}^\lambda(T;Y),\hat{I}_{\mathcal{D}_\text{OPT}}^\lambda(X;T))$ is shown in the top panel for one random split of the data set. Furthermore, the corresponding values of the test mutual informations for the hyperparameters tested by the FST procedure, along with the pair $(I^{\lambda^*}(T;Y), I^{\lambda^*}(X;T))$ finally returned by IB-MHT are depicted in the bottom panel.

\begin{figure}[t]
    \centering
    \includegraphics[width=\columnwidth]{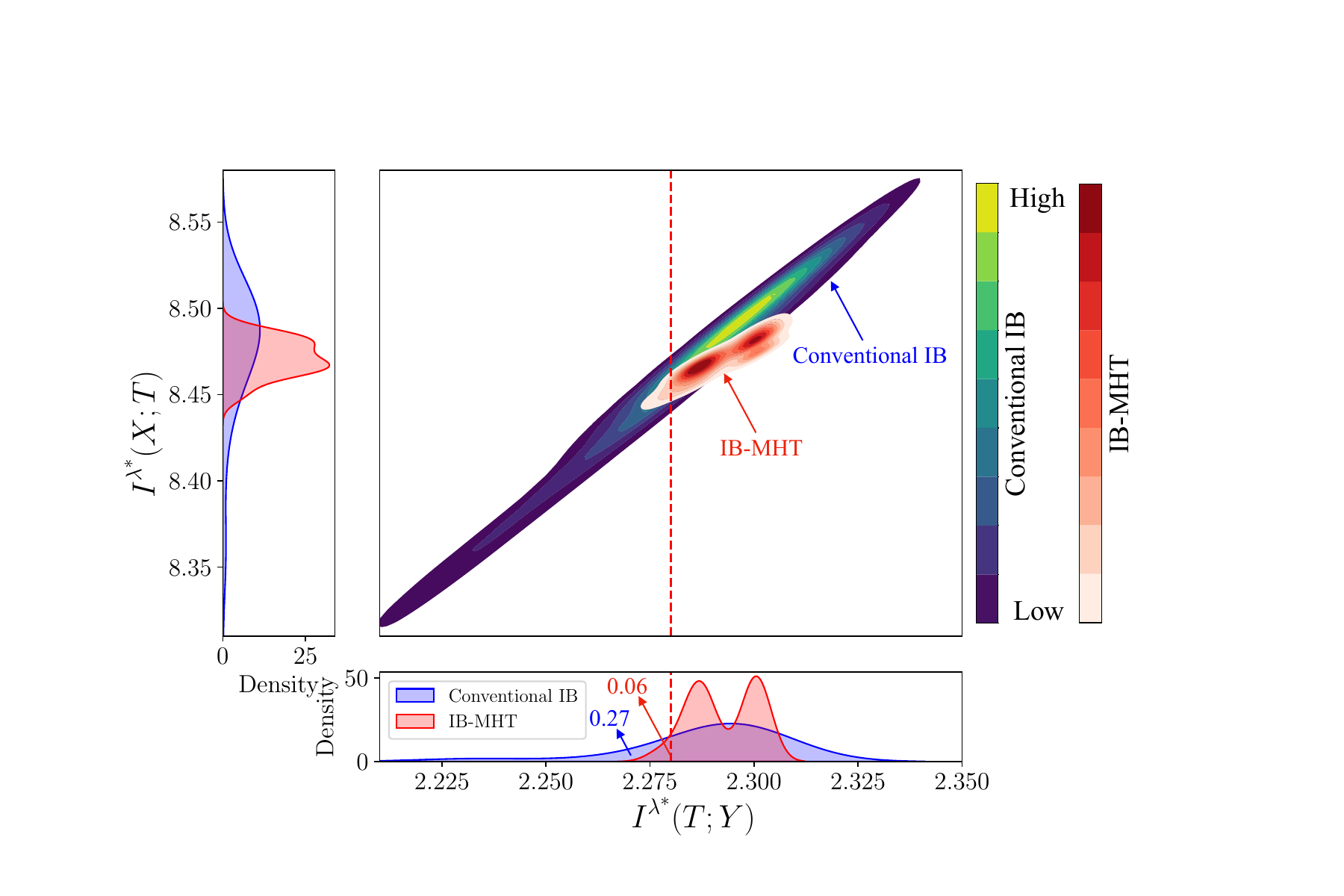}
    \caption{Joint distributions of the mutual informations $I^{\lambda^*}(T;Y)$ and $I^{\lambda^*}(X;T)$ obtained by using a conventional IB solver (Section \ref{sec:conventional_IB}) and IB-MHT for the classical IB problem (\ref{eq:classic_IB}) using 50 trials of Algorithm \ref{alg:Pareto}. The outage probability for conventional IB and IB-MHT are reported to be 0.27 and 0.06, respectively.}
    \label{fig:VIB}
\end{figure}

Considering now 50 independent splits $(\mathcal{D}_{\text{OPT}}, \mathcal{D}_{\text{MHT}})$ of the calibration data set $\mathcal{D}$, Fig. \ref{fig:VIB} shows the joint distribution for the mutual informations $I^{\lambda^*}(T;Y)$ and $I^{\lambda^*}(X;T)$ estimated on the test set, alongside the corresponding marginal distributions for the conventional IB solution reviewed in Section \ref{sec:conventional_IB} and for IB-MHT. 

IB-MHT is observed to satisfy the requirement $I^\lambda(T;Y)\geq 2.28$ with an outage level below the target $\delta = 0.1$, thus meeting the constraint (\ref{eq:relaxed_constraint}). In contrast, conventional IB violates the requirement (\ref{eq:relaxed_constraint}), returning hyperparameter $\lambda^*$ with mutual information $I^{\lambda^*}(T;Y)<2.28$ for a fraction 0.27 of the cases. IB-MHT is also observed to have a significantly lower variability in terms of the obtained pair $(I^{\lambda^*}(T;Y), I^{\lambda^*}(X;T))$. Furthermore, despite failing to satisfy the requirement (\ref{eq:relaxed_constraint}), conventional IB returns objective values $I^{\lambda^*}(X;T)$ for problem (\ref{eq:IB}) with mean 8.46 and a standard deviation as high as 0.05. IB-MHT can instead guarantee the requirement (\ref{eq:relaxed_constraint}), while also yielding objectives $I^{\lambda^*}(X;T)$ with comparable mean 8.47 and significantly smaller standard deviation 0.01.

\subsection{Deterministic IB Problem}

We now consider the deterministic IB problem (\ref{eq:deterministic_IB2}) \cite{strouse2017deterministic}. We form the set of candidate hyperparameters as $\Lambda=\Gamma \times B$, where $\Gamma$ and $B$ consist of 10 logarithmically spaced points in the intervals $[10^{-3}, 1]$ and $[10^{-4}, 1]$, respectively. The values of $\alpha$ and $\delta$ are set to $2.28$ and $0.1$, respectively.

\begin{figure}[t]
    \centering
    \includegraphics[width=\columnwidth]{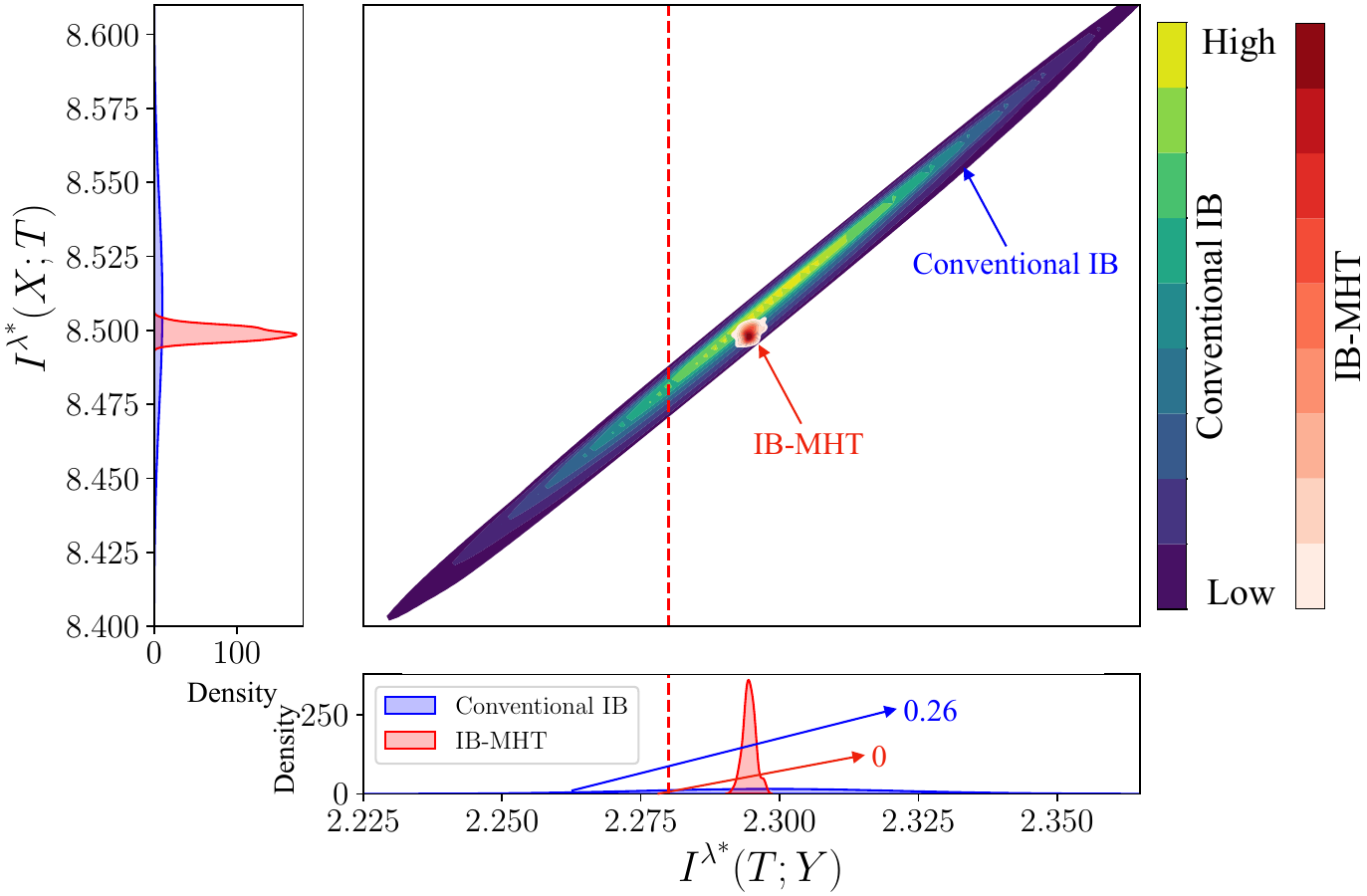}
    \caption{Joint distributions of the mutual informations $I^{\lambda^*}(T;Y)$ and $I^{\lambda^*}(X;T)$ obtained by using a conventional IB solver (Section \ref{sec:conventional_IB}) and IB-MHT for the deterministic IB problem (\ref{eq:deterministic_IB2}) using 50 trials of Algorithm \ref{alg:Pareto}. The outage probability for conventional IB and IB-MHT are reported to be 0.26 and near zero, respectively.}
    \label{fig:det}
\end{figure}

As in Fig. \ref{fig:VIB}, Fig. \ref{fig:det} shows the joint and marginal distributions of the obtained mutual informations $(I^{\lambda^*}(T;Y),I^{\lambda^*}(X;T))$ for conventional IB and IB-MHT. The general conclusions are aligned with Fig. \ref{fig:VIB}. Moreover, the gains of IB-MHT are seen to be more pronounced than for the classical IB problem (\ref{eq:classic_IB}), owing to the larger number of hyperparameters to be optimized. Notably, unlike conventional IB, IB-MHT can leverage the larger number of hyperparameters to ensure a greater control over the requirement (\ref{eq:relaxed_constraint}), which is met here with probability of outage near zero. The standard deviation on the attained objective is also decreased from 0.01 to 0.002 as compared to conventional IB, while preserving a similar mean value.

Additionally, IB-MHT is shown to offer significant improvements in terms of the stability of the distillation process. The variability in the mutual information terms $I(S;T)$ and $I(S;X)$ is markedly reduced when using IB-MHT. For example, the standard deviation of $I(S;X)$ decreases from 0.04 in conventional distillation to 0.01 when using IB-MHT. This indicates that IB-MHT provides better control over the trade-off between retaining relevant information and minimizing irrelevant details, resulting in a more robust and reliable distillation process.

\section{Experiments for Knowledge Distillation in Text Representation}

In this section, we evaluate the performance of IB-MHT when applied to the the IB Knowledge Distillation (IBKD) method proposed in \cite{zhang2023text}. Via KD, a smaller, more efficient student language model is trained to mimic the behavior of a larger, more powerful teacher model.

IBKD leverages the IB principle to control the flow of information from the teacher to the student. To elaborate, define as $X$ the input text, as $Y$ the representation of the text produced by the teacher model, and as $T$ the text representation output by the student model. The model representation is extracted from the last layer of the language model \cite{zhang2023text}. IBKD seeks to compress the student's representation $T$, while ensuring that it retains task-relevant information about the representation $Y$ produced by the teacher, filtering out unnecessary details from the input data $X$.

The IB problem is addressed in \cite{zhang2023text} in the modified form
\begin{equation}
\label{eq:IBKD}
    \underset{P_{T|X}}{\text{minimize}} \quad -I(T;Y) + \lambda I(X;T),
\end{equation} 
where the hyperparameter $\lambda>0$ multiplies the compression term $I(X;T)$.

In \cite{zhang2023text}, hyperparameter $\lambda$ is set as $\lambda=1$ for all simulations. Here, we apply the proposed IB-MHT to problem (\ref{eq:IBKD}) in order to guarantee statistical reliability of the student model’s performance via hyperparameter optimization. To this end, we set $\alpha = 17.8$ and $\delta = 0.1$ in the statistical constraint (\ref{eq:relaxed_constraint}). In our simulations, we use as a benchmark the conventional setting $\lambda = 1$. As in the previous section, we split the data into two subsets, $\mathcal{D}_{\text{OPT}}$ and $\mathcal{D}_{\text{MHT}}$, and performed 50 independent trials of IB-MHT. The candidate hyperparameter space $\Lambda$ consists of 100 linearly spaced candidate values in the interval $[0.01,2]$.
\begin{figure}[t]
    \centering
    \includegraphics[width=\columnwidth]{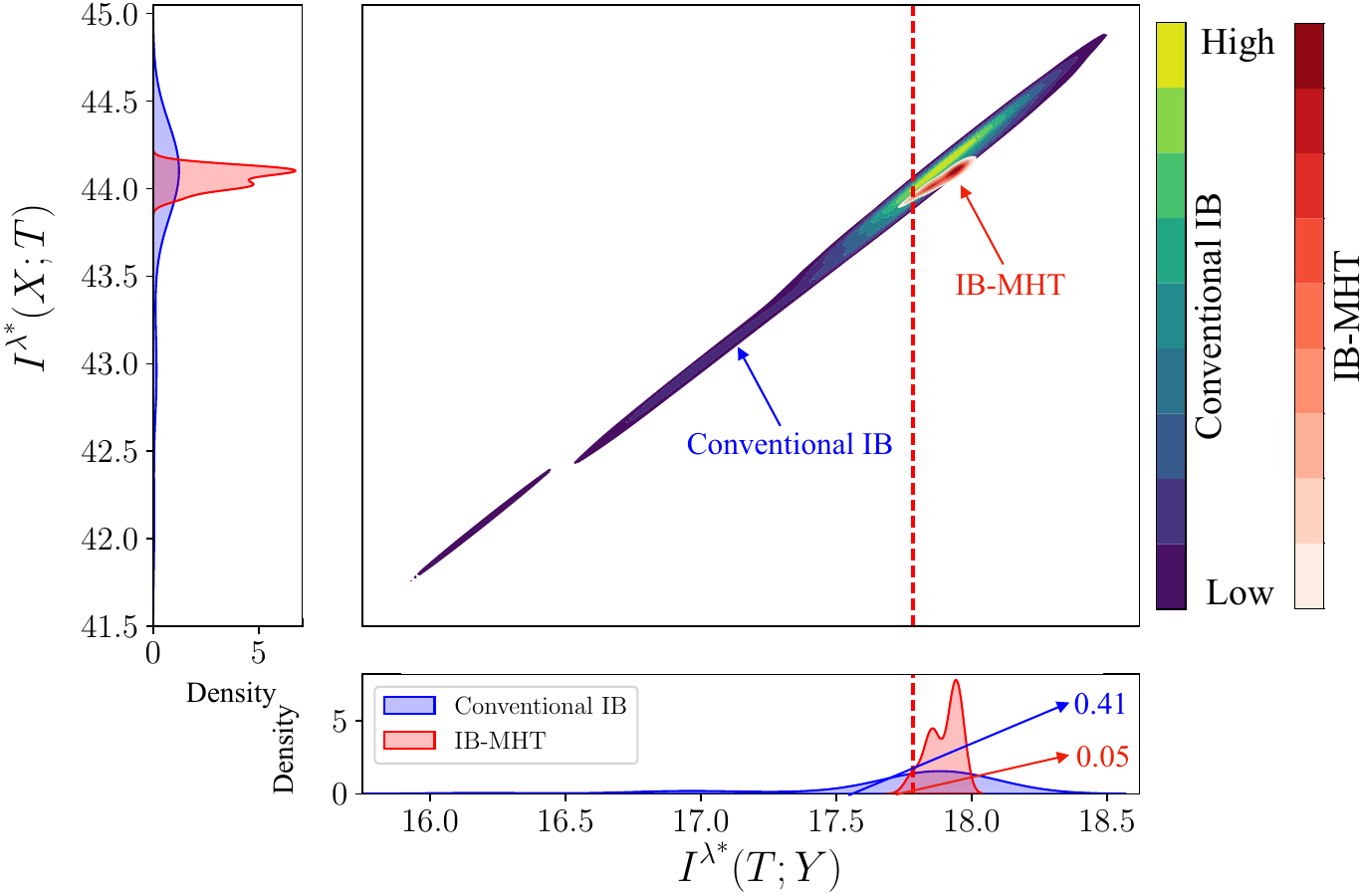}
    \caption{Joint distributions of the mutual informations $I^{\lambda^*}(T;Y)$ and $I^{\lambda^*}(X;T)$ obtained by using a fixed setting $\lambda = 1$ and IB-MHT for the IBKD optimization problem (\ref{eq:IBKD}) using SimCSE-RoBERTa\(_{\text{large}}\) and TinyBERT as the teacher and student models, respectively, and 50 trials of Algorithm \ref{alg:Pareto}. The outage probability for conventional IB and IB-MHT are reported to be 0.41 and 0.05, respectively.}
    \label{fig:tiny}
\end{figure}

Following \cite{zhang2023text}, we adopt the Semantic Textual Similarity (STS) dataset, which is a standard benchmark for knowledge distillation tasks. The subsets $\mathcal{D}_\text{OPT}$ and $\mathcal{D}_\text{MHT}$ contain 2,874 and 2,875 examples, respectively. Furthermore, SimCSE-RoBERTa\(_{\text{large}}\)\footnote{\url{https://huggingface.co/princeton-nlp/sup-simcse-roberta-large}} is used as the teacher model, while the student model is TinyBERT\footnote{\url{https://huggingface.co/nreimers/TinyBERT_L-4_H-312_v2}} or MiniLM\footnote{\url{https://huggingface.co/nreimers/MiniLM-L6-H384-uncased}}.

\begin{figure}[t]
    \centering
    \includegraphics[width=\columnwidth]{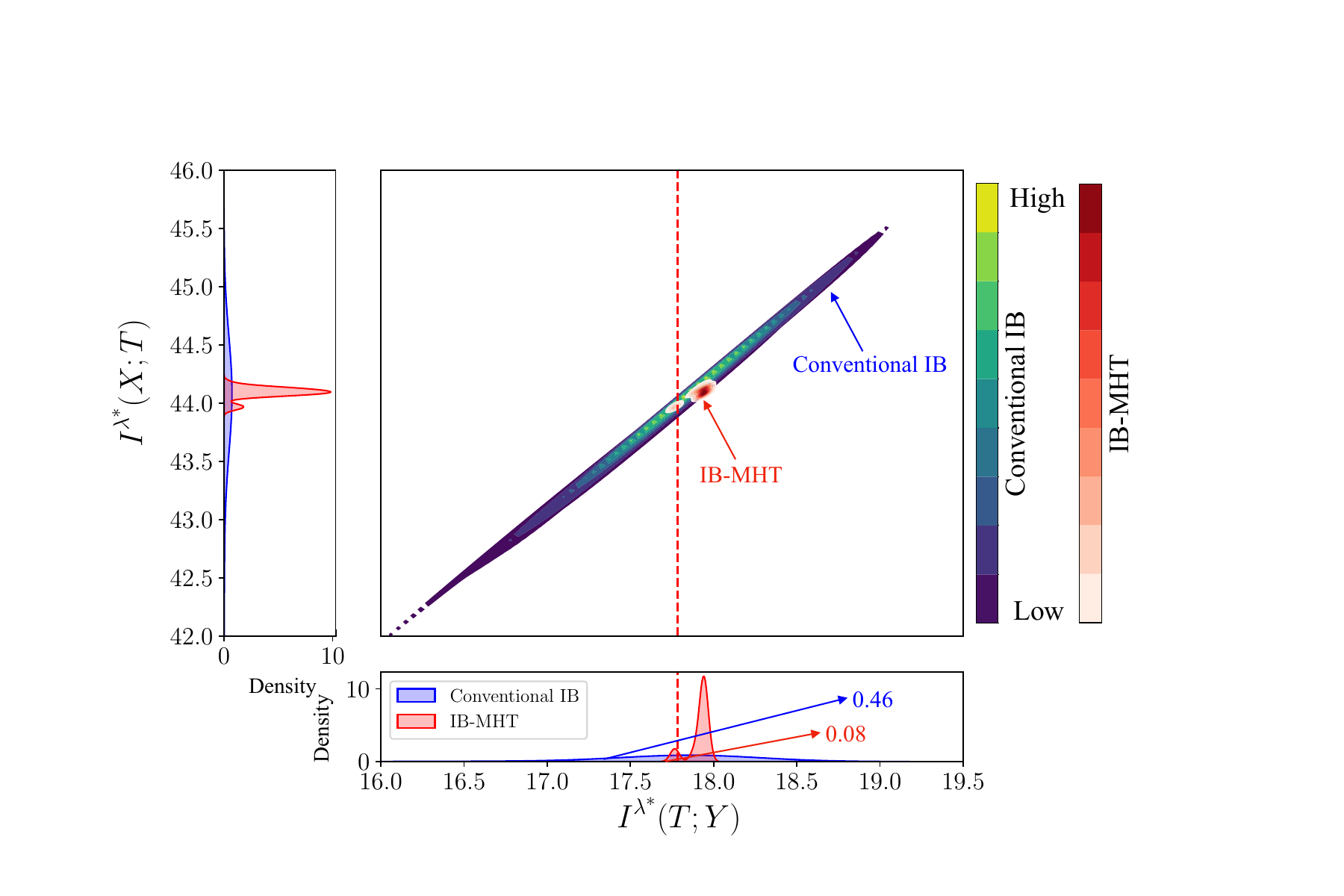}
    \caption{Joint distributions of the mutual informations $I^{\lambda^*}(T;Y)$ and $I^{\lambda^*}(X;T)$ obtained by using a fixed setting $\lambda = 1$ and IB-MHT for the IBKD optimization problem (\ref{eq:IBKD}) using SimCSE-RoBERTa\(_{\text{large}}\) and MiniLM as the teacher and student models, respectively, and 50 trials of Algorithm \ref{alg:Pareto}. The outage probability for conventional IB and IB-MHT are reported to be 0.46 and 0.08, respectively.}
    \label{fig:mini}
\end{figure}

The simulation results for TinyBERT and MiniLM are illustrated in Fig. \ref{fig:tiny} and Fig. \ref{fig:mini}, respectively. Both figures show that IB-MHT successfully guarantees the mutual information constraint (\ref{eq:relaxed_constraint}) on $I(T;Y)$, ensuring that the student model provably retains sufficient information from the teacher. Specifically, IB-MHT satisfies the constraint $I(T;Y) \geq \alpha = 17.8$ with an outage probability below the target level of $\delta = 0.1$ for both student models. In contrast, the conventional setting $\lambda=1$ violates this constraint in approximately $41\%$ and $46\%$ of the cases for TinyBert and MiniLM, respectively.

Additionally, Fig. \ref{fig:tiny} and Fig. \ref{fig:mini} show that IB-MHT offers significant improvements in terms of the stability of the distillation process. In particular, the variability in the mutual information terms $I(X;T)$ and $I(T;Y)$ is markedly reduced when using IB-MHT. For example, the standard deviation of the mutual information $I(T;Y)$ for MiniLM decreases from 1.05 for the conventional setting $\lambda = 1$ to 0.05 when using IB-MHT.

\begin{figure}[t]
    \centering
    \includegraphics[width=\columnwidth]{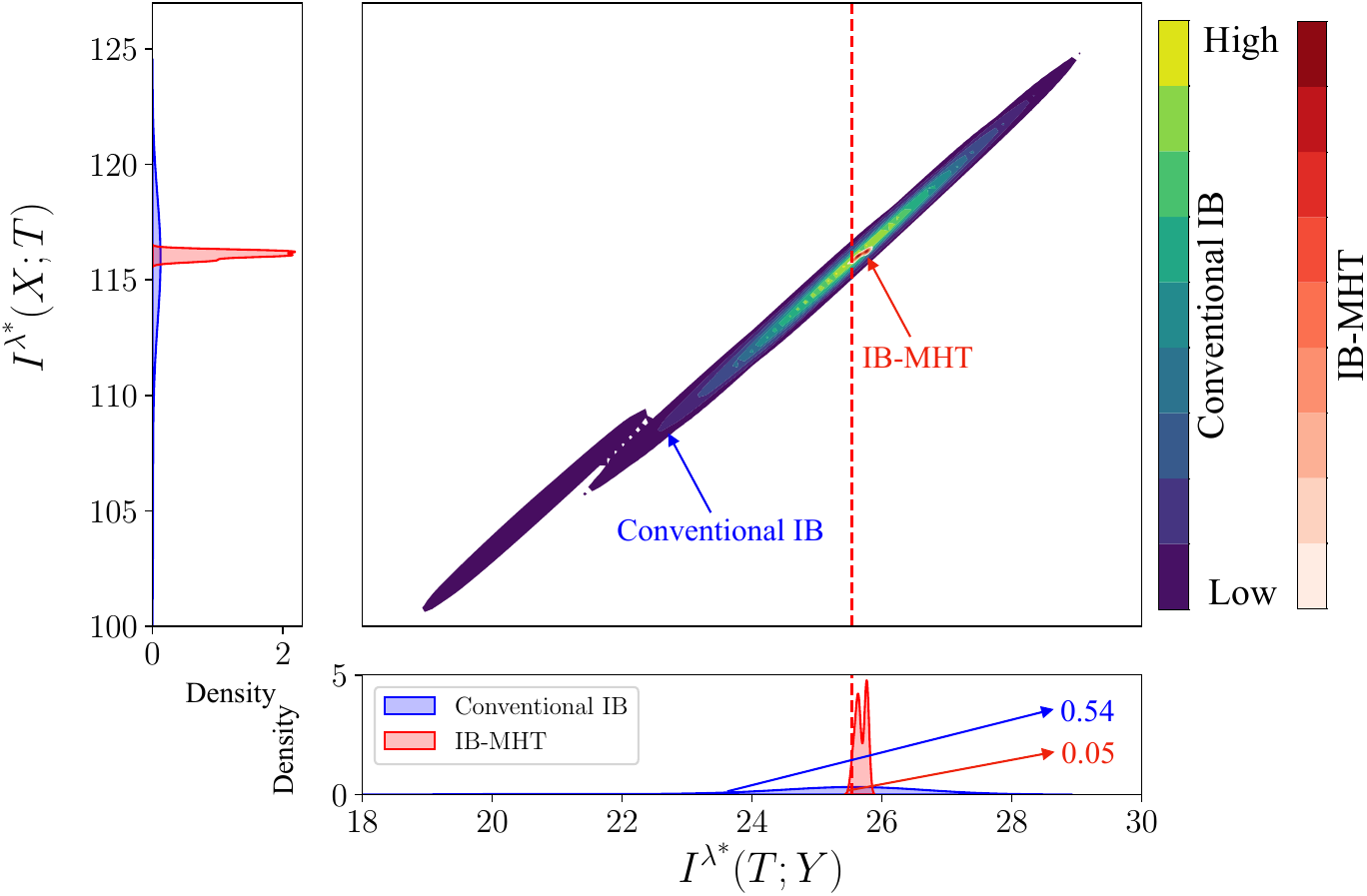}
    \caption{Joint distributions of the mutual informations $I^{\lambda^*}(T;Y)$ and $I^{\lambda^*}(X;T)$ obtained by using a fixed setting $\lambda = 1$ and IB-MHT for the IBKD optimization problem (\ref{eq:IBKD}) using CoCondenser and TinyBERT as the teacher and student models, respectively, and 50 trials of Algorithm \ref{alg:Pareto}. The outage probability for conventional IB and IB-MHT are reported to be 0.54 and 0.05, respectively.}
    \label{fig:tiny2}
\end{figure}

\begin{figure}[t]
    \centering
    \includegraphics[width=\columnwidth]{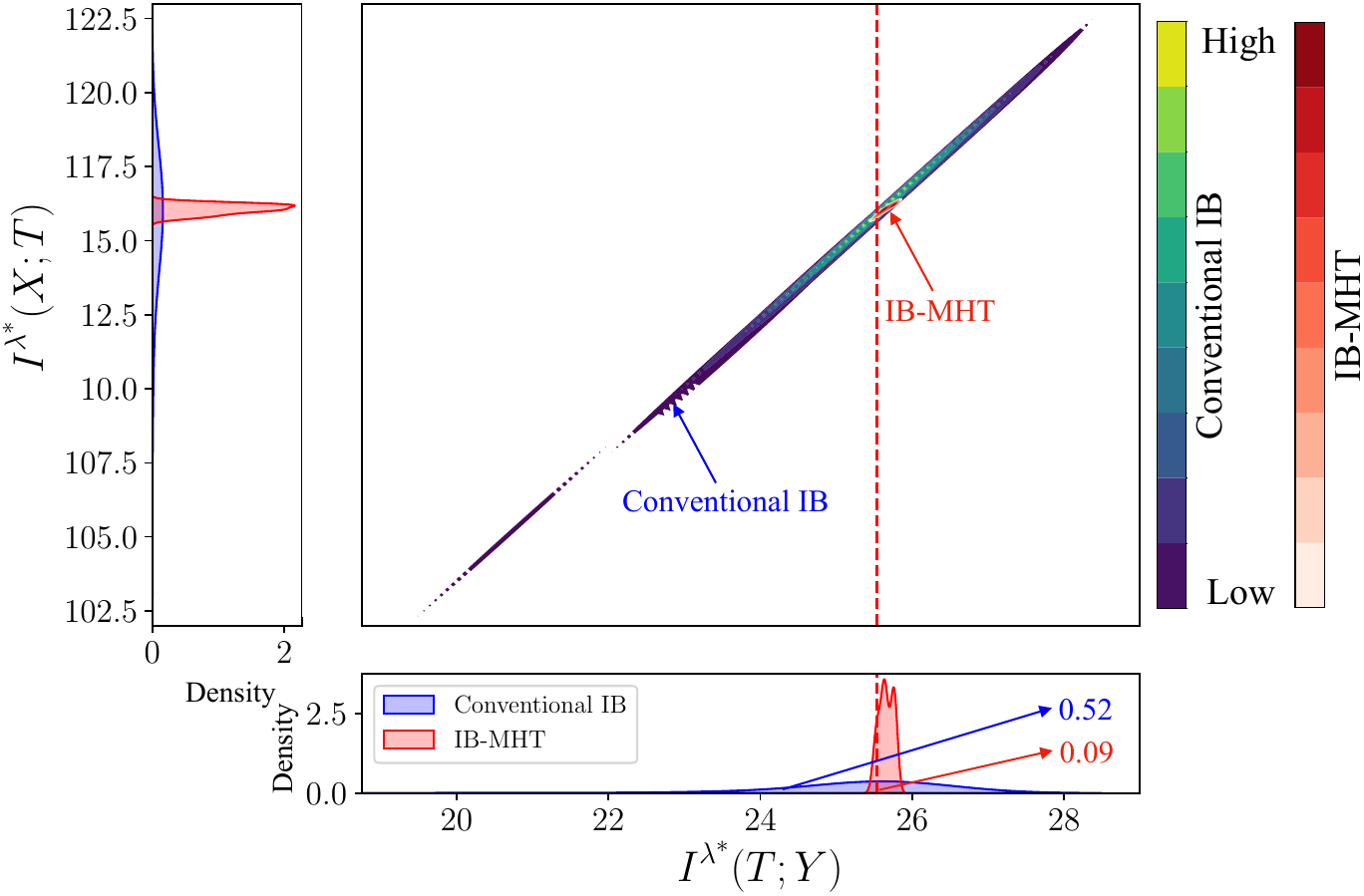}
    \caption{Joint distributions of the mutual informations $I^{\lambda^*}(T;Y)$ and $I^{\lambda^*}(X;T)$ obtained by using a fixed setting $\lambda = 1$ and IB-MHT for the IBKD optimization problem (\ref{eq:IBKD}) using CoCondenser and MiniLM as the teacher and student models, respectively, and 50 trials of Algorithm \ref{alg:Pareto}. The outage probability for conventional IB and IB-MHT are reported to be 0.52 and 0.09, respectively.}
    \label{fig:mini2}
\end{figure}

We also performed experiments on the MS MARCO passage dataset \cite{bajaj2016ms}, using CoCondenser\footnote{\url{https://huggingface.co/Luyu/co-condenser-marco-retriever}} as the teacher model. Fig. \ref{fig:tiny2} and Fig. \ref{fig:mini2} illustrate the results with TinyBERT and MiniLM as the student models, respectively. In both cases, IB-MHT consistently provides the desired statistical guarantees, outperforming the conventional setting $\lambda = 1$. Overall, this section demonstrates that IB-MHT continues to offer robust performance and statistical validity even on more complex, real-world datasets.

\section{Conclusion}
\label{sec:conclusion}

In this paper, we have proposed IB-MHT, a statistically valid approach for solving the information bottleneck problem using hyperparameter optimization. Unlike conventional methods that rely on heuristic hyperparameter tuning, IB-MHT leverages multiple hypothesis testing (MHT), wrapping around existing IB solvers to ensure statistical guarantees on the mutual information constraints. Our experimental results on both classical and deterministic IB formulations demonstrated the benefit of IB-MHT, including for advanced applications such as text distillation in language models. These results demonstrate IB-MHT's versatility and effectiveness in handling modern, state-of-the-art tasks, further validating its applicability beyond standard IB formulations. Future research could explore the extension of IB-MHT to continuous variables, as well as applying similar techniques to other information-theoretic metrics such as convex divergences.

\section*{Acknowledgements}
This work was supported by the European Union’s Horizon Europe project CENTRIC (101096379), by the Open Fellowships of the EPSRC (EP/W024101/1), and by the EPSRC project (EP/X011852/1).


\begin{thebibliography}{10}

\bibitem{tishby2000information}
N.~Tishby, F.~Pereira, and W.~Bialek, ``The information bottleneck method,'' in {\em Proc. Allerton Conference on Communication, Control and Computation}, 2001.


\bibitem{zaidi2020information}
A.~Zaidi, I.~Estella-Aguerri, and S.~Shamai, ``On the information bottleneck problems: Models, connections, applications and information theoretic views,'' {\em Entropy}, vol.~22, no.~2, p.~151, 2020.

\bibitem{slonim1999agglomerative}
N.~Slonim and N.~Tishby, ``Agglomerative information bottleneck,'' in {\em Proc. Advances in Neural Information Processing Systems}, 1999.

\bibitem{achille2018information}
A.~Achille and S.~Soatto, ``{Information dropout: Learning optimal representations through noisy computation},'' {\em IEEE Transactions on Pattern Analysis and Machine Intelligence}, vol.~40, no.~12, pp.~2897--2905, 2018.

\bibitem{higgins2017beta}
I.~Higgins, L.~Matthey, A.~Pal, C.~Burgess, X.~Glorot, M.~Botvinick, S.~Mohamed, and A.~Lerchner, ``beta-{VAE}: Learning basic visual concepts with a constrained variational framework,'' in {\em Proc. International Conference on Learning Representations}, 2017.

\bibitem{strouse2017deterministic}
D.~Strouse and D.~J. Schwab, ``The deterministic information bottleneck,'' {\em Neural Computation}, vol.~29, no.~6, pp.~1611--1630, 2017.

\bibitem{alemi2016deep}
A.~A. Alemi, I.~Fischer, J.~V. Dillon, and K.~Murphy, ``Deep variational information bottleneck,'' in {\em Proc. International Conference on Learning Representations}, 2017.

\bibitem{tishby2015deep}
N.~Tishby and N.~Zaslavsky, ``Deep learning and the information bottleneck principle,'' in {\em Proc. IEEE Information Theory Workshop (ITW)}, 2015.

\bibitem{laufer2022efficiently}
B.~Laufer-Goldshtein, A.~Fisch, R.~Barzilay, and T.~S. Jaakkola, ``{Efficiently Controlling Multiple Risks with Pareto Testing},'' in {\em Proc. International Conference on Learning Representations}, 2023.

\balance

\bibitem{angelopoulos2021learn}
A.~N. Angelopoulos, S.~Bates, E.~J. Cand{\`e}s, M.~I. Jordan, and L.~Lei, ``{Learn then test: Calibrating predictive algorithms to achieve risk control},'' {\em arXiv preprint arXiv:2110.01052}, 2021.

\bibitem{shwartz2017opening}
R.~Shwartz-Ziv and N.~Tishby, ``Opening the black box of deep neural networks via information,'' {\em arXiv preprint arXiv:1703.00810}, 2017.

\bibitem{stefani2014confidence}
A.~G. Stefani, J.~B. Huber, C.~Jardin, and H.~Sticht, ``Confidence intervals for the mutual information,'' {\em International Journal of Machine Intelligence and Sensory Signal Processing}, vol.~1, no.~3, pp.~201--214, 2014.

\bibitem{rice2007mathematical}
J.~Rice, {\em Mathematical Statistics and Data Analysis}.
\newblock Cengage Learning, 2007.

\bibitem{bauer1991multiple}
P.~Bauer, ``Multiple testing in clinical trials,'' {\em Statistics in Medicine}, vol.~10, no.~6, pp.~871--890, 1991.

\bibitem{zhang2023text}
Y.~Zhang, D.~Long, Z.~Li, and P.~Xie, ``Text representation distillation via information bottleneck principle,'' \emph{arXiv preprint arXiv:2311.05472}, 2023.

\bibitem{bajaj2016ms}
P.~Bajaj, D.~Campos, N.~Craswell, L.~Deng, J.~Gao, X.~Liu, R.~Majumder, A.~McNamara, B.~Mitra, T.~Nguyen, \emph{et al.},  
``MS MARCO: A human generated machine reading comprehension dataset,'' {\em arXiv preprint arXiv:1611.09268}, 2016.





\end{thebibliography}
\end{document}